\documentclass[pre,showpacs,showkeys,preprintnumbers,amsmath,amssymb,superscriptaddress,twocolumn]{revtex4-1}
\usepackage[english]{babel}
\usepackage{makeidx} 
\usepackage{graphicx} 
\usepackage{dcolumn} 
\usepackage{array} 
\usepackage{amssymb} 
\usepackage{amsmath}
\usepackage{textcomp}
\usepackage{multirow}
\usepackage{subfigure}
\usepackage{eucal}
\usepackage{mathrsfs}
\usepackage[all]{xy}
\usepackage{epstopdf}

\usepackage{color}

\usepackage{float} 
\usepackage{amsmath} 
\usepackage{amsfonts} 
\usepackage{bm} 

\begin{document}
\bibliographystyle{unsrtnat}

\title{Non-Arrhenius behavior and fragile-to-strong transition of glass-forming liquids}

\author{A. C. P. Rosa Jr.}\affiliation{Grupo de Informa\c{c}\~{a}o Qu\^{a}ntica e F\'{i}sica Estat\'{i}stica, Centro de Ci\^{e}ncias Exatas e das Tecnologias, Universidade Federal do Oeste da Bahia. Rua Bertioga, 892, Morada Nobre I, 47810-059 Barreiras, Bahia, Brazil.}
\author{C. Cruz\email{clebson.cruz@ufob.edu.br}} \affiliation{Grupo de Informa\c{c}\~{a}o Qu\^{a}ntica e F\'{i}sica Estat\'{i}stica, Centro de Ci\^{e}ncias Exatas e das Tecnologias, Universidade Federal do Oeste da Bahia. Rua Bertioga, 892, Morada Nobre I, 47810-059 Barreiras, Bahia, Brazil.}\email{clebson.cruz@ufob.edu.br}
\author{W. S. Santana}\affiliation{Grupo de Informa\c{c}\~{a}o Qu\^{a}ntica e F\'{i}sica Estat\'{i}stica, Centro de Ci\^{e}ncias Exatas e das Tecnologias, Universidade Federal do Oeste da Bahia. Rua Bertioga, 892, Morada Nobre I, 47810-059 Barreiras, Bahia, Brazil.}
\author{E. Brito}\affiliation{Grupo de Informa\c{c}\~{a}o Qu\^{a}ntica e F\'{i}sica Estat\'{i}stica, Centro de Ci\^{e}ncias Exatas e das Tecnologias, Universidade Federal do Oeste da Bahia. Rua Bertioga, 892, Morada Nobre I, 47810-059 Barreiras, Bahia, Brazil.}
\author{M. A. Moret}\affiliation{Programa de Modelagem Computacional - SENAI - CIMATEC, 41650-010 Salvador, Bahia, Brazil}\affiliation{Universidade do Estado da Bahia - UNEB, 41150-000 Salvador, Bahia, Brazil}
\date{\today}

\begin{abstract}
Characterization of the non-Arrhenius behavior of glass-forming liquids is a broad avenue for research toward the understanding of the formation mechanisms of noncrystalline materials. In this context, this paper explores the main properties of the viscosity of glass-forming systems, considering super-Arrhenius diffusive processes. We establish the viscous activation energy as a function of the temperature, measure the degree of fragility of the system,  and characterize the fragile-to-strong transition through the {standard Angell's plot.} Our results show that the non-Arrhenius behavior observed in fragile liquids can be understood through the non-Markovian dynamics that characterize the diffusive processes of these systems. Moreover, the fragile-to-strong transition corresponds to  {a change in the spatio-temporal range of} correlations during the glass transition process.
\end{abstract}
\maketitle

\section{Introduction}

The study of viscosity in glass-forming liquids has received considerable attention in glass manufacturing and fundamental research  in physics \cite{Zheng2016,mauro2014grand,mauro2014glass,PhysRevE.100.022139,Ikeda2019,Aymen2015,jaccani2017modified,peng2016decoupled,yildirim2016revealing}. Glasses are basically noncrystalline materials, but they have some structural order on a microscopic scale. This order is conditioned by their chemical composition and the thermal history of the system, which led the material to the glass transition process \cite{Zheng2016,mauro2014grand,mauro2014glass}. In this regard, the characterization of its viscosity as a function of temperature, during the cooling process, provides relevant information about the structural properties of the forming material, because it is associated with the local interactions between moving molecules and their immediate neighborhood \cite{Zheng2016,mauro2014grand,mauro2014glass}.

In an earlier work by our group \cite{PhysRevE.100.022139}, a nonadditive stochastical model was developed for the non-Arrhenius behavior of diffusivity and viscosity in supercooled fluids, close to the glass transition. This model consists of a nonhomogeneous continuity equation that corresponds to a class of nonlinear Fokker--Planck equations \cite{Frank2005,Ribeiro2013,Casas2019}, whose drag and diffusive coefficients are associated with anomalous diffusion processes, and its stationary solutions maximize nonadditive {entropies \cite{PhysRevE.100.022139,Schwmmle2009,dosSantosMendes2017}}. We also provided  a reliable measurement of the fragility index \cite{PhysRevE.100.022139,Angell2002,Zheng2016} and an estimation of the fragile-to-strong {transition \cite{PhysRevE.100.022139,Shi2018,Ikeda2019,Lucas2019}} through the ratio $T_t/T_g$, where $T_g$ is the glass transition temperature and $T_t$ is the threshold temperature for the super-Arrhenius process.  

Therefore, by utilizing a nonadditive stochastic model, one can understand how the non-Arrhenius behavior of viscosity is associated with the existence of {disordered short-range local structures \cite{Tanaka2000,Shi2018}}, arising from local interactions between moving molecules and their neighborhood during the glass transition process \cite{PhysRevE.100.022139}{. This} explains why nonequilibrium viscosity is a function not only of temperature, but also of thermal history and chemical composition of the forming material, since nonlinear forms of the Fokker--Planck equation are related to non-Markovian stochastic dynamics \cite{PhysRevE.100.022139,Frank2005}. 

In this paper, we explore the main properties of the viscosity of glass-forming liquids through its super-Arrhenius behavior \cite{PhysRevE.100.022139}. Following the Angell's definitions \cite{Zheng2016,Angell2002}, we analyze the temperature dependence of viscosity and recalculated the fragility index corresponding to the viscous activation energy measured by the glass transition temperature. We establish the relationship between the fragility index and the characteristic exponent from which we classify the different non-Arrhenius difussive processes, providing a new indicator of the degree of fragility in supercooled liquids. Moreover, we use a two-state model \cite{Tanaka2000,Shi2018} in order to {simulate the variation of the activation energy in the fragile-to-strong transition,} observed in water-like systems \cite{Shi2018,Lucas2019} and metallic glass-forming liquids \cite{Ikeda2019,Gallino2017}. 

Our results show that the fragile-to-strong transition can be understood as the {change in the spatio-temporal range of  correlations} as the glass-forming system reaches the glass-transition temperature. As a consequence, the standard Arrhenius behavior corresponds to the short-range spatio-temporal correlations in the dynamic properties of the system, associated with linear forms of the Fokker--Planck equation. Likewise, the non-Arrhenius behavior observed in fragile liquids can be understood as a consequence of the non-Markovian dynamics that characterize these diffusive processes. Therefore, this work provides to the literature a path toward the physical interpretation of the fragile-to-strong transition, observed in supercooled liquids. 

\section{Viscosity of glass-forming liquids}

The viscosity of glass-forming liquids can be defined as the inverse of the generalized mobility of the fluid, such that for the stationary state of the system \cite{PhysRevE.100.022139}, its temperature dependence can be written as
\begin{equation}
\eta (T)=\eta_{\infty }\left[1-(2-m)\frac{E}{k_B T}\right]^{-\gamma}~,
\label{eq1}
\end{equation}
where $\eta_\infty$ is the viscosity in the high-temperature limit \cite{PhysRevE.100.022139}; $E$ is a generalized energy related to the molar energy in a reaction-diffusion model \cite{PhysRevE.100.022139}; $k_B$ is the Boltzmann constant; and $\gamma=(m-1)/(2-m)$ is a characteristic exponent, with $m$ being a generalized parameter which classifies different non-Arrhenius {behaviors \cite{Truhlar2001,PhysRevE.100.022139}}. The condition {$m=2$ implies} an Arrhenius standard plot,where the viscosity of the supercooled liquid increases exponentially as cooling approaches the glass transition temperature; if $m>2$, Eq. (\ref{eq1}) describes the sub-Arrhenius process, associated with the nonlocal quantum effects on the dynamics of the system \cite{Cavalli2014,CarvalhoSilva2016,MeanaPaeda2011,Silva2013};  
otherwise, if $m<2$, the viscosity shows a super-Arrhenius behavior, indicating that classical transport phenomena dominate the diffusion process \cite{Silva2013,CarvalhoSilva2016}, and the viscosity, Eq. (\ref{eq1}), diverges when the temperature reaches the threshold value $T_t=(2-m)E/k_B$. 

Owing to the high variation of the viscosity during the glassforming process, it is usual to represent its measurements in a logarithmic scale. Thus, we can rewrite Eq. (\ref{eq1}) for the super-Arrhenius case as
\begin{equation}
log_{10} \eta = \gamma \sum_{n=1}^{\infty}\frac{1	}{n} \left(\frac{T_t}{T}\right)^n + log_{10} \eta_\infty~,
\label{eq2}
\end{equation}
where the sum corresponds to the power series expansion of the logarithmic function. Therefore, the high-temperature regime $(T >>T_t)$ allows a first-order approximation in Eq. (\ref{eq2}), describing an Arrhenius-like behavior even though $ m\neq2$. These higher-order approximations correspond to some empirical expressions proposed in the literature to fit viscosity experimental data as a function of temperature \cite{Aymen2015,Valantina2015}.

In a thermally activated viscous process, the activation energy corresponds to a potential barrier associated with resistance to molecular motion due to the neighborhood action \cite{Zheng2016}. From Eq.(\ref{eq1}), the temperature dependence of the viscous activation energy is given by
\begin{equation}
E_{vis}(T) =  \frac{(m-1)E}{1-(m-1) \frac{E}{\gamma k_B T}}~.
\label{eq3}
\end{equation}

As can be seen from the above equation, $E_{vis} = (m-1)E_A$, where $E_A$ is the measurement of the activation energy measurement, whose thermal behavior has been illustrated at reference \onlinecite{PhysRevE.100.022139}. The viscous activation energy, Eq. (\ref{eq3}), is an increasing function with respect to the reciprocal temperature for the super-Arrhenius processes $(m<2)$, decreasing for sub-Arrhenius processes $(m>2)$, and temperature-independent for the condition $m=2$, in which the generalized energy $E$, in Eq. (\ref{eq3}), becomes the Arrhenius standard activation energy. Furthermore, $E_{vis} \rightarrow \infty $ for $T \rightarrow T_t$ in the super-Arrhenius processes reflects the viscosity divergence for the threshold temperature.

\section{Fragility}

Glass-forming liquids are classified into two categories \textit{fragile} and  \textit{strong} \cite{Zheng2016}, depending on the viscosity dependence with temperature. A system is considered \textit{strong} if its behavior is close to the standard Arrhenius behavior, where the viscosity presents a linear dependence with the reciprocal temperature due to the temperature-independent behavior of the activation energy \cite{PhysRevE.100.022139}, {and \textit{fragile}} if they exhibit super-Arrhenius behavior, where the activation energy increases with the reciprocal temperature \cite{PhysRevE.100.022139,angell1988perspective,martinez2001thermodynamic}.   

One method of characterizing the fragility index of glass-forming systems is through the standard Angell's plot \cite{Zheng2016,Angell1988,Angell2002}, which consists of the logarithmic dependence of viscosity as a function of the ratio $T_g /T$. Therefore,  from Eq. (\ref{eq1}), we can obtain\begin{equation}
log_{10} \eta = -\gamma log_{10} \left[\frac{1-\frac{T_t}{T}}{1-\frac{T_t}{T_g}}\right] + 12~,
\label{eq4}
\end{equation}
where we consider the reference value $10^{12}$ Pa.s as the viscosity at the glass transition temperature \cite{PhysRevE.100.022139}. Therefore, the slope of the Angell's plot for $T =T_g$ defines the fragility index of the liquid \cite{Zheng2016,Angell1988,Angell2002}. From Eq.(\ref{eq4}), we can obtain this index as
\begin{equation}
M_\eta^{(A)} = \gamma \frac{T_t / T_g}{1-T_t / T_g}~.
\label{eq5}
\end{equation}
The smaller the difference between the glass transition temperature ($T_g$) and the threshold temperature ($T_t$), the more fragile the glass-forming system will be. Moreover, as can be seen from Eq. (\ref{eq3}), the fragility index corresponds to the measurement of the viscous activation energy at the glass transition temperature can be expressed as
\begin{equation}
M_\eta^{(A)} = \frac{E_{vis} (T_g)}{k_BT_g}~.
\label{eq6}
\end{equation}

From this theoretical model, one can uniquely relate the fragility index to the proposed $\gamma$ exponent as
\begin{equation}
M_\eta^{(A)} = \gamma \left( 10^{B/\gamma} -1\right)~,
\label{eq7}
\end{equation}
where $B = 12 - log \eta_\infty$.  

\begin{figure}[htp]
	\centering
	{\includegraphics[scale=0.43]{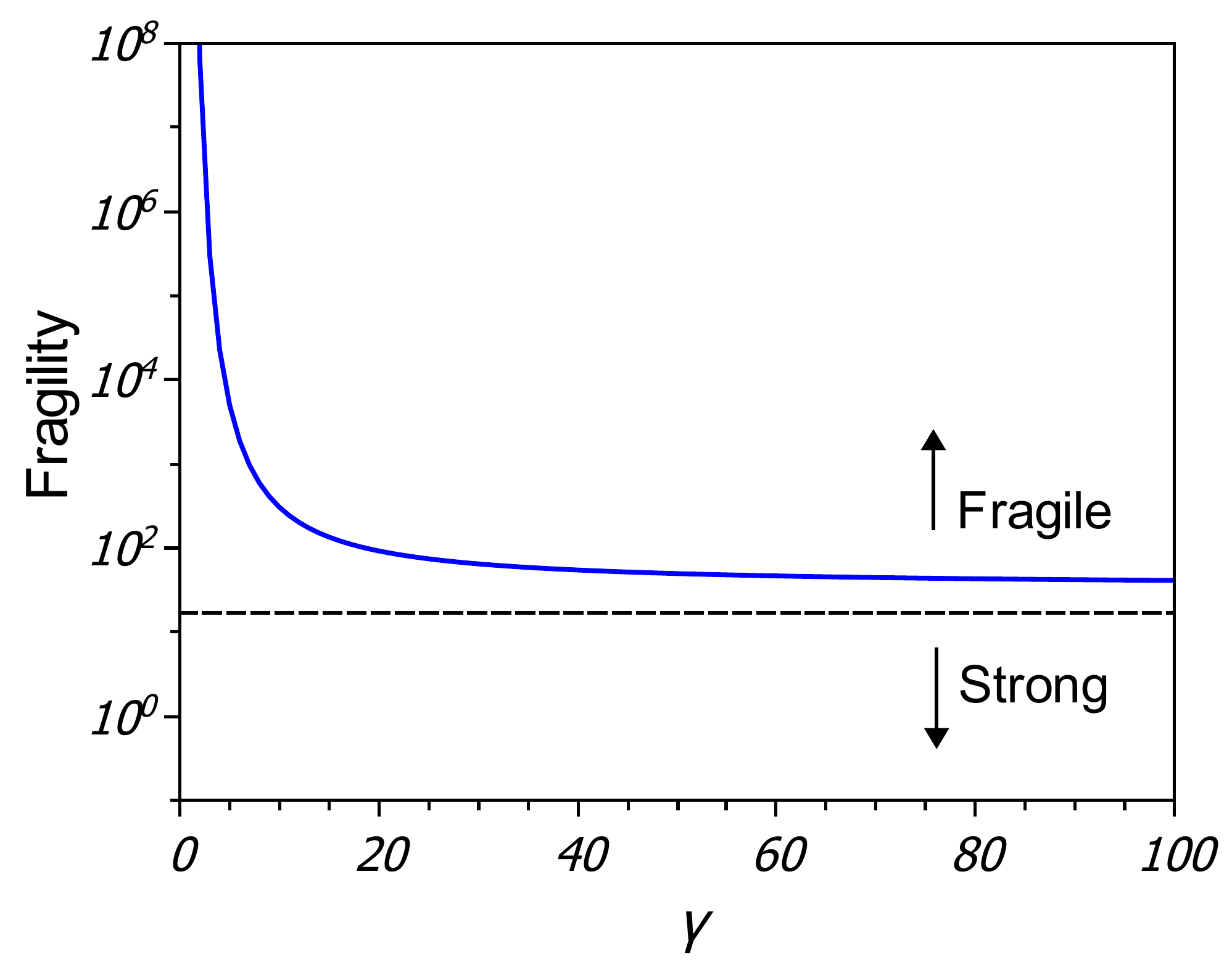}}\\
	\caption{(Color online) The fragility index as a function of the exponent $\gamma$. The curve ({solid blue line}) corresponds to Eq.(6) for $B=15$. The horizontal line ({dashed black line}) corresponds to the value of $M_\eta^{(A)} = 15$ and is an asymptotic limit between fragile and strong behaviors.}
	\label{fig:fig1}
\end{figure}

Figure \ref{fig:fig1} shows the fragility index $M_\eta^{(A)}$ as a function of the exponent $\gamma$. As can be seen, this exponent represents a reliable resolution for quantifying the fragility degree of a glass-forming system.
Experimental evidences indicate a universal behavior for high-temperature viscosity \cite{Mauro2009}, such that $\eta_\infty \approx 10^{-3}$Pa.s, which implies $B \approx 15$. According to Eq. (\ref{eq7}), the fragility index has an asymptotic limit $ M_\eta^{(A)} = 15 $ (dashed line) for the condition $\gamma \rightarrow \infty$, which is the threshold between fragile and strong regimes. Therefore, the glass-forming system is classified as strong only at $M_\eta^{(A)} \rightarrow 15$. In this limit, its viscosity follows Arrhenius's standard behavior. However, fragile systems obey the condition $M_\eta^{(A)} > 15$ and the Eq. (\ref{eq1}) describes the viscosity dependence of the temperature at super-Arrhenius viscous processes. 

\begin{figure}[htp]
	\centering
	{\includegraphics[scale=0.43]{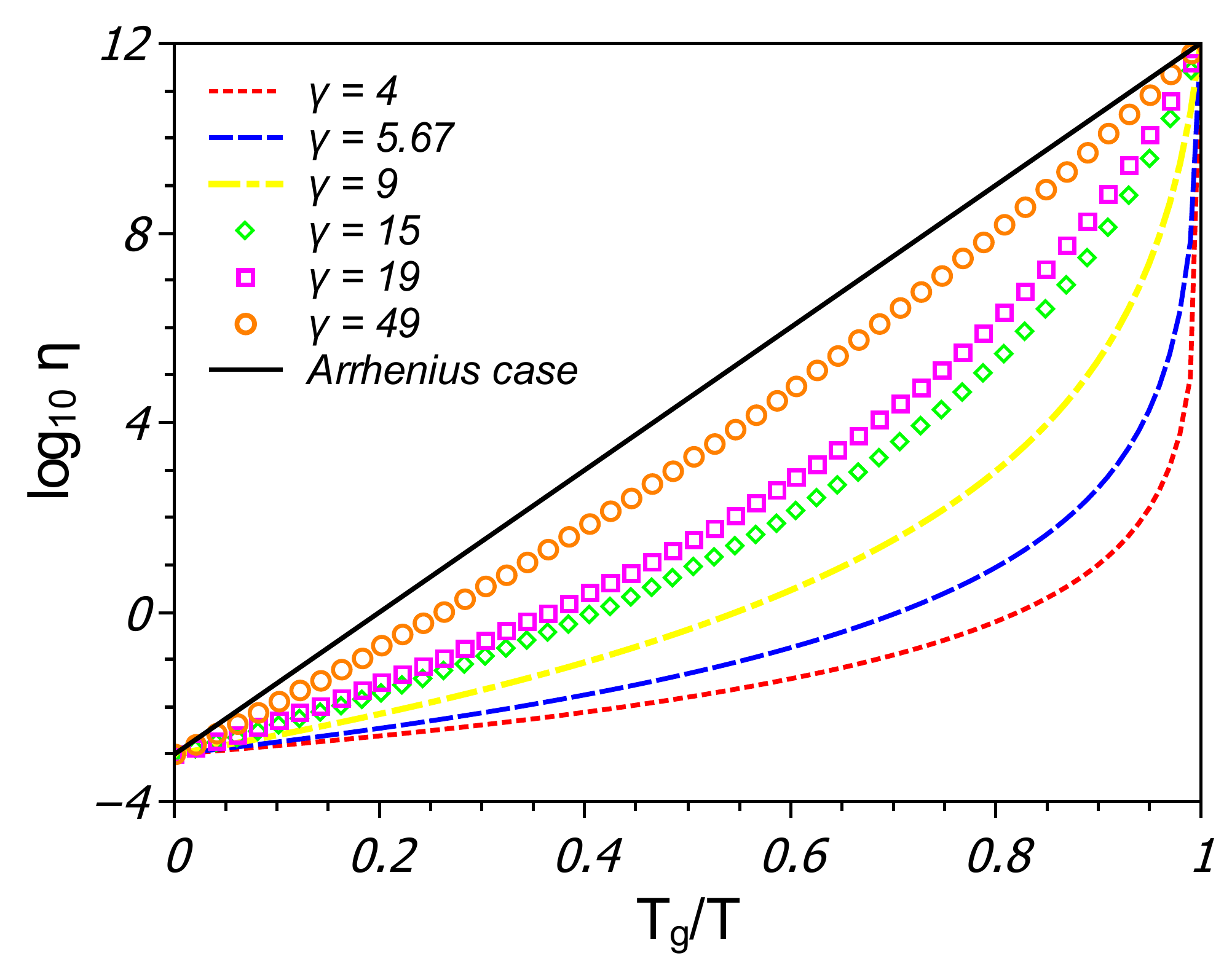}}\\
	\caption{(Color online) Angell's plot of the non-Arrhenius behavior of diffusivity and viscosity in supercooled fluids, obtained from our nonadditive stochastic model. Eq. (\ref{eq4}) was plotted  as a function of $T_g$ scaled by the reciprocal temperature for different $\gamma$ values. The plotted dashed lines represent super-Arrhenius liquids with different levels of fragility, whereas the solid black line corresponds to the typical Arrhenius behavior of strong liquids.}
	\label{fig:fig2}
\end{figure}

Moreover, from Eq. (\ref{eq7}), we determine the ratio $ T_t / T_g $ for different values of $\gamma$ to obtain the Angell's plot through the Eq. (\ref{eq4}). Figure \ref{fig:fig2} shows the standard Angell's plot obtained from a nonadditive stochastical model for the viscosity in supercooled fluids, it  reproduces the pattern displayed by the original Angell' graphic \cite{Angell1988}. We establish a relationship between the fragility index, as defined by Angell, and the characteristic exponent $\gamma$ of our nonadditive stochastical model. Therefore, the $\gamma$ exponent can also provides {an indicative} of the fragility index of glass-forming liquids. In this context, the fragility corresponds to a measurement of {the change in the} spatio-temporal correlations on the dynamic properties of these systems throughout the glass transition process, which is equivalent to assessing the dependence of the viscosity with the thermal history and chemical composition of the system. 

\section{Fragile-to-strong transition}

Some glass-forming systems do not follow the fragility curves shown in Angell's plot and undergo a transition between the fragile and strong behaviors as the system approaches the glass transition temperature, leading to a crossover between the super-Arrhenius {behavior and} the standard Arrhenius behavior (solid {black line in Figure \ref{fig:fig2}}). Both types of behaviors have distinct activation energies as the system approaches the glass transition temperature. This phenomenon is known as {fragile-to-strong} transition \cite{Shi2018,Lucas2019}.

The theoretical basis of the {fragile-to-strong} transition in supercooled liquids has been unclear in glass science; hence, it has been the subject of study in several experimental and phenomenological works in the past few years \cite{Lucas2019,Ikeda2019,Shi2018,Zhang2010,Zhou2015,Wei2015,Hajime2018,Sukhomlinov2019}. 
Because this transition corresponds to the shift in viscosity measurements between the super-Arrhenius and standard Arrhenius behaviors for viscosity measurement, as the system cools \cite{Shi2018,Lucas2019}, it is not possible to classify the glass-forming system as either fragile or strong liquid. However, the physical nature of the fragile-to-strong transition can be explained by our {nonadditive stochastic model, since the} $ T_t / T_g $ ratio in Eq. (\ref{eq4}) acts as an indicator of the fragile-to-strong transition attributable to this model because its variation changes the fragility index and the slope rate of the curves in the standard Angell's plot. This implies that the generalized energy $E$ in Eq. (\ref{eq1}) is no longer constant, as a result of the changes in the dynamic properties of the supercooled liquid produced by the transition.

In this regard, we calculate the viscosity in a system undergoing a fragile-to-strong transition by applying the two-state model \cite{Tanaka2000,Rui2018}, which has shown promising results in characterizing the dynamic properties of supercooled water \cite{Shi2018}. {It is worth noting that the two-state model is not the microscopic correspondent of our nonadditive stochastic model, we present a macroscopic model for the dynamics of supercooled liquids without the construction of a corresponding microscopic model. Thus, we take the two-state model \cite{Tanaka2000,Rui2018}, only to simulate the variation in the generalized energies $E$}. In this  {two-state model}, the supercooled liquid is a dynamic mixture of two distinct liquid states, which we labeled as state I and state II, corresponding to different local structures at the molecular levels. The proportion of the liquids in the system is a function of the pressure and temperature \cite{Shi2018,Rui2018}. For temperatures farther from $T_g$, state I will dominate the system, while state II is dominant near the glass transition temperature. Therefore, we define the generalized energy $E$ as
\begin{equation}
E = E^{(I)} + \left(E^{(II)}-E^{(I)}\right)s_I
\label{eq8}
\end{equation}
where $E^{(I)}$ and $E^{(II)}$ are the generalized energies for state I and II respectively, and $s_I$ is the fraction of state I whose dependence on temperature is empirically characterized by the Boltzmann factor \cite{Shi2018}. 

\begin{figure}[htp]
	\centering
	{\includegraphics[scale=0.45]{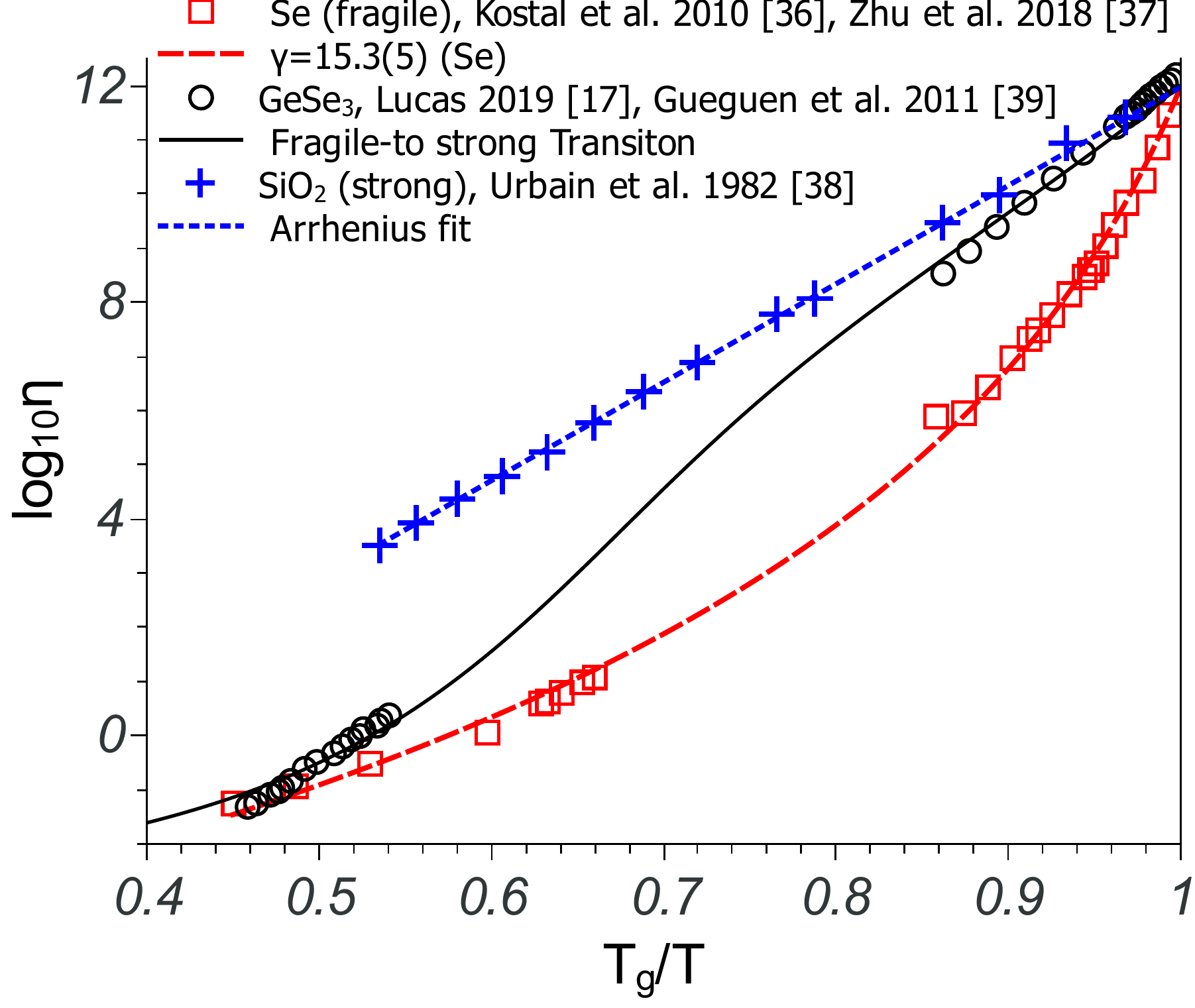}}\\
	\caption{(Color online) {Angell plot of experimental results obtained in the literature  \cite{Lucas2019,Kotl2010,Zhu2018,Urbain1982,Gueguen2011} and the respective fits obtained from our nonadditive stochastic model. Dashed red line describes the fit of the super-Arrhenius behavior (fragile liquid) presented by the selenium melt (red open squares) \cite{Kotl2010,Zhu2018}; the dotted blue line represents the linear fit of the standard Arrhenius behavior (strong liquid) observed for liquid silica, SiO$_{2}$, (blue crosses) \cite{Urbain1982}; {solid black line} shows the fragile-to-strong transition calculated for the GeSe$_{3}$ glass-forming system in comparison with its viscosity measurements reported in the literature (black open circles) \cite{Lucas2019,Gueguen2011}.}}
	\label{fig:fig3}
\end{figure}

Figure \ref{fig:fig3} shows the transition curve ({solid black line}) on the Angell's plot; this curve is bounded by the super-Arrhenius behavior ({dotted blue line}) on the lower side and by the Arrhenius standard behavior ({dashed red line}) on the upper side. {In order to show the applicability of our model, we fit a set of experimental data of glass-forming systems selected in the literature \cite{Lucas2019,Kotl2010,Zhu2018,Urbain1982,Gueguen2011}.} 

{As can be seen, we fit $\gamma$ coefficient for the super-Arrhenius behavior (fragile liquid) presented by the selenium melt \cite{Kotl2010,Zhu2018}, where we found $\gamma=15.3(5)$. In contrast, for liquid silica (SiO$_{2}$) \cite{Urbain1982}, a linear fit is a sufficient condition to achieve its standard Arrhenius behavior (strong liquid). Thus, rewriting Eq. (\ref{eq4}) at the limit that $\gamma$ goes to infinity in  Eq. (\ref{eq1}) we found the viscous activation energy $E=0.51(5)$ MJ/mol for the liquid silica, while the value reported in the literature is $0.515$ MJ/mol \cite{Urbain1982}. Moreover, we simulate the fragile-to-strong transition of the glass forming system GeSe$_{3}$, based on its viscosity measurements reported in the literature \cite{Lucas2019,Gueguen2011}. The transition curve (solid black line), was obtained using the two-state model for the variation of the generalized energy, Eq. (\ref{eq8}), where we estimate the generalized exponent $\gamma$ in each state, for which $\gamma= 4.05$ is associated with the fragile regime (state I), whereas $\gamma= 37.69$ is associated with a strong liquid regime (state II).}

{According to the two-state model, in the high-temperature regime, disordered short-range local structures dominate over the supercooled liquid; such condition allows the association of state I to a fragile liquid. On the other hand, state II is characterized by the presence of ordered short-range local structures, which causes the supercooled liquid to follow standard Arrhenius behavior near $T_g$. In this context, the spatio-temporal correlations in the dynamic properties of the system are different for each state. Due to this fact, the non-Arrhenius behavior observed in fragile liquids can be understood in terms of the non-Markovian dynamics that characterize these diffusive processes as a natural condition of our nonadditive stochastic model.} Therefore, the physical nature of the fragile-to-strong transition can be interpreted as {a change in the spatio-temporal range of} correlations as the glass-forming system attains the glass-transition temperature.

\section{Conclusion}

In summary, our main result was to provide to the literature the theoretical basis of the physical interpretation of the fragile-to-strong transition in supercooled liquids. Through a nonadditive stochastic model, we characterized the most relevant properties associated with the measurement of the viscosity in glass-forming systems. 
We established the fragility index, $M_\eta ^{(A)}$, in terms of the characteristic exponent, $\gamma$, of our nonadditive stochastic model, providing an alternative method to characterize the degree of fragility of the glass-forming system. Finally, {we show the applicability of our model fitting a set of viscosity measurements for glass-forming systems selected in the literature. The fragile-to-strong transition was calculated for the GeSe$_{3}$ glass-forming system in the Angell's plot,} by varying the generalized energies through the two-state model and determining the viscosity of the super-Arrhenius behavior in the high-temperature regime as well as the standard Arrhenius behavior near the glass transition temperature. {Thus, we show that our model describes the fragile-to-strong transition and the non-Arrhenius behavior of glass-forming liquids, based on experimental results reported in the literature}.

Therefore, the non-Arrhenius behavior observed in the viscosity of fragile liquids is shown to be a consequence of the non-Markovian dynamics that characterize the diffusive processes of these systems, and the fragile-to-strong transition, observed in water-like systems and metallic glass-forming liquids, can be understood as the {change in the spatio-temporal range of} correlations during the glass transition process. Thus, a nonadditive stochastic model provides the theoretical basis for a consistent physical interpretation of the dynamic properties of the glass-forming systems, opening a broad avenue for research in glass science toward the understanding of the formation mechanisms of noncrystalline materials.

\begin{acknowledgments}
This study was financed in part by the CNPq and  the \textit{Coordena\c{c}\~{a}o de Aperfei\c{c}oamento de Pessoal de N\'{i}vel Superior - Brasil} (CAPES) - Finance Code 001.
\end{acknowledgments}

\end{document}